\def\be{\begin{equation}}
\def\ee{\end{equation}}
\def\bea{\begin{eqnarray}}
\def\eea{\end{eqnarray}}
\def\av#1{\langle {#1} \rangle}
\def\rr{{\vec{r}}}
\def\rp{{\vec{r}^{\,\prime}}}
\def\RR{{\vec{R}}}
\def\Rp{{\vec{R}^\prime}}

\documentclass[preprint,showpacs,superscriptaddress,floatfix,aps,prl]{revtex4}
\usepackage{graphics}

\begin{document}

\title{Spatial field correlation, the building block of mesoscopic fluctuations}
\date{\today}

\author{P. Sebbah}
\affiliation{Laboratoire de Physique de la Mati{\`{e}}re
Condens{\'{e}}e, Universit{\'{e}} de Nice - Sophia Antipolis,
\\Parc Valrose, 06108, Nice Cedex 02, France}

\author{B. Hu}
\affiliation{Department of Physics, Queens College of the City
University of New York, Flushing, New York 11367}

\author{A. Z. Genack}
\affiliation{Department of Physics, Queens College of the City
University of New York, Flushing, New York 11367}

\author{R. Pnini}
\affiliation{Department of Physics, Technion-Israel Institute of
Technology, Haifa 32000, Israel}

\author{B. Shapiro}
\affiliation{Department of Physics, Technion-Israel Institute of
Technology, Haifa 32000, Israel}

\begin{abstract}
The absence of self averaging in mesoscopic systems is a
consequence of long-range intensity correlation. Microwave
measurements suggest and diagrammatic calculations confirm that
the correlation function of the normalized intensity with
displacement of the source and detector, $\Delta R$ and $\Delta
r$, respectively, can be expressed as the sum of three terms, with
distinctive spatial dependences. Each term involves only the sum
or the product of the square of the field correlation function, $F
\equiv F_{E}^2$. The leading-order term is the product, the next
term is proportional to the sum. The third term is proportional to
$[F(\Delta R)F(\Delta r) + [F(\Delta R)+F(\Delta r)] + 1]$.
\end{abstract}


\maketitle

Short-range correlation in waves transmitted through random media
is manifest in the intensity speckle pattern. The leading
contribution, $C_1$, to the cumulant correlation function $C$ of
intensity normalized to its ensemble average on the output surface
of the sample is given by the square of the field correlation
function, $C_1=F_E^2$. Neglecting internal reflection from the
surface, its dependence upon displacement $\Delta r$ on the output
surface is given by $C_1(\Delta r) = F_E^2(\Delta r) \equiv
F(\Delta r) = (\sin k \Delta r/ k \Delta r)^2 \exp(-\Delta
r/\ell_s)$, where $k$ is the wave vector and $\ell_s$ is the
scattering mean free path \cite{Shapiro}. This term dominates
intensity fluctuations. Defining a correlation length, $\delta r
$, as the first zero of $C_1$ gives $\delta r = \pi/k=\lambda/2$.
The intensity is correlated far beyond $\delta r$ as a result of
scattering within the medium
\cite{Stephen,Feng88,mello,GenackPRL90,kk,dutch} so that intensity
values in remote speckle spots are not statistically independent.
This gives rise to two additional contributions to $C$, which can
therefore be expressed as, $C=C_1 + C_2 + C_3$
\cite{Feng88,Shapiro99}, and leads to greatly enhanced mesoscopic
fluctuations \cite{AltshulerAltshuler}. The $C_2$ term produces an
enhancement in total transmission fluctuations over that given by
the field factorization approximation by a factor of $L/\ell$,
where $\ell$ is the transport mean free path and $L$ is the sample
length \cite{Stephen,dutch}. The $C_3$ term is the source of
universal conductance fluctuations, which are enhanced by a factor
of $(L/\ell)^2$ \cite{AltshulerAltshuler,Laibowitz}. The magnitude
of $C_1$ at a point is unity as a result of normalization, whereas
the magnitudes of $C_2$ and $C_3$ are expansions in $1/g$ with
leading terms of order $1/g$ and $1/g^2$ \cite{Feng88},
respectively, where $g$ is the dimensionless conductance. Since
the onset of localization is at $g = 1$ \cite{gang4}, the two
terms beyond the field factorization approximation for $C$,
reflect the approach to localization.

In this Letter, we use microwave measurements and diagrammatic
calculations to show that each of the contributions to $C$ may be
expressed in terms of the square of the field correlation function
with regard to displacements of the source, $\Delta R$ and
detector, $\Delta r$. The $C_1$ term is $F(\Delta R)F(\Delta r)$,
the $C_2$ term is proportional to $[F(\Delta R) + F(\Delta r)]$,
while the $C_3$ term is proportional to $[F(\Delta R)F(\Delta r) +
[F(\Delta R) + F(\Delta r)] + 1 ]$. When the intensity correlation
is considered at a shifted frequency, $\Delta\nu$, the full
correlation function remains a sum of three terms each being a
product of the corresponding terms in C, thus, $C_i =
A_i(\Delta\nu) C_i(\Delta r,\Delta R)\  (i=1,2,3)$. Absorption
alters the magnitudes of $C_2$ and $C_3$, but it does not change
the spatial structure of these terms.

Initial measurements of angular intensity correlation, carried out
in the far field of weakly scattering media, gave $C$, which was
essentially equal to $C_1$ \cite{Freund88,Li94}. Recently,
measurements of the spatial correlation of the field on the sample
surface have yielded the $C_1$ contribution directly \cite{PRE00}.
Measurements of intensity correlation between points on the sample
surface and the interior of the sample on a scale greater than the
wavelength have allowed the observation of $C_2$
\cite{GenackPRL90}. In addition measurements have been made of the
spectral correlation of the $C_1$ \cite{Genack87} and $C_2$ terms
\cite{Garcia89,Ad90,GGPS} and of the temporal correlation of the
$C_1$ \cite{Maret87}, $C_2$ \cite{FSPRB97} and $C_3$
\cite{FSPRL98} terms in colloidal samples. However, the variation
of $C$ on a subwavelength scale, as well as the structure of
correlation with displacement of both the source and detector,
have not been reported previously. The present measurements and
calculations allow us to discern the structure of intensity
correlation and to relate it to the correlation of the underlying
field.

Measurements are made in a disordered dielectric sample contained
within a reflecting tube. Radiation of frequency $\nu$ emitted by
a source antenna at $\vec{R}$ at one end of the tube is detected
at a point $\vec{r}$ at the other end. We denote the intensity at
$\vec{r}$ due to a source at $\vec{R}$ by
$I_{\nu}(\vec{r},\vec{R})$ and consider the normalized cumulant
correlation function,
\be 
C(\Delta r, \Delta R ) \equiv \av{\delta I_\nu (\rr, \RR ) \delta
I_{\nu^\prime} (\rp, \Rp)}/ \av{I_{\nu}(\rr,
\RR)}\av{I_{\nu^\prime}(\rp, \Rp)} \ , \label{cor} \ee where
$\delta I_{\nu}$ is the deviation of the intensity from its
ensemble average value, $\Delta r = |\rr - \rp |$ and $\Delta R =
|\RR - \Rp|$ are the displacements across the output and input
surfaces, respectively, and $\av{\ldots}$ denotes the average over
an ensemble of random realizations. The leading contribution to
$C$ obtained by factorizing the fields is
\cite{Stephen,Shapiro,PRE00,FreundEliyahou}:
\be  
C_1 (\Delta r, \Delta R ) = |\av{ E_{\nu}(\rr, \RR)
E^\ast_{\nu^\prime} (\rp, \Rp)}|^2/ \av{I_{\nu}(\rr, \RR)} \av{
I_{\nu^\prime}(\rp, \Rp)} \ . \label{factor} \ee

The samples studied are random mixtures of 1.27-cm-diameter
polystyrene spheres at a volume filling fraction of 0.52. They are
contained inside a 100 cm long copper tube with a diameter of 7.5
cm capped with thin Plexiglas face plates. Field and intensity
spectra are obtained in the frequency range 16.8 - 17.8 GHz with a
step size of 625 kHz using a vector network analyzer. Measurements
are taken in an ensemble of 690 random samples by rotating the
tube to create new configurations of the spheres after each set of
spectra are taken. Field spectra are obtained in each sample
realization by translating an antenna detector along a line to
each of 50 locations separated by 1.06 mm on the output surface
for each of two fixed antenna sources at the incident surface,
which are separated by $\Delta R = 3 \pm 0.1 \ {\rm cm}\equiv d$.
Field spectra are also taken by translating the source along a
line for each of two fixed antenna detectors at the output
surface, separated by $\Delta r=d$. At the separation $d$, the
short-range term has largely decayed.
The antennas are aligned perpendicular to the line of translation.
Intensity spectra are obtained by squaring the field spectra.

\begin{figure}[h!]
\scalebox{.7}{\includegraphics{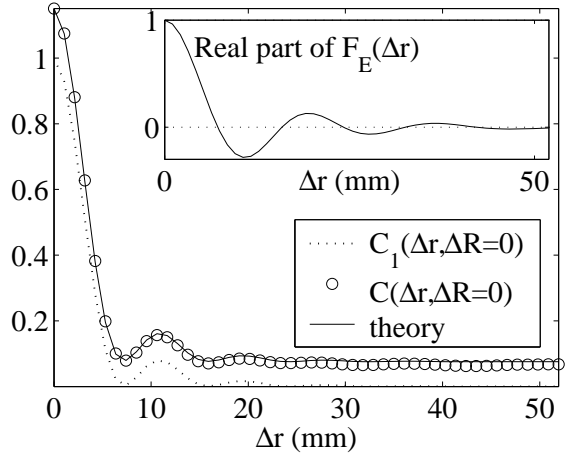}} \caption{Plots of
$C(\Delta r, \Delta R\!=\!0)$ and $C_1(\Delta r, \Delta R\!=\!0)$,
and theoretical fit to $C$.} \label{fig1}
\end{figure}
\begin{figure}[h!]
\scalebox{.7}{\includegraphics{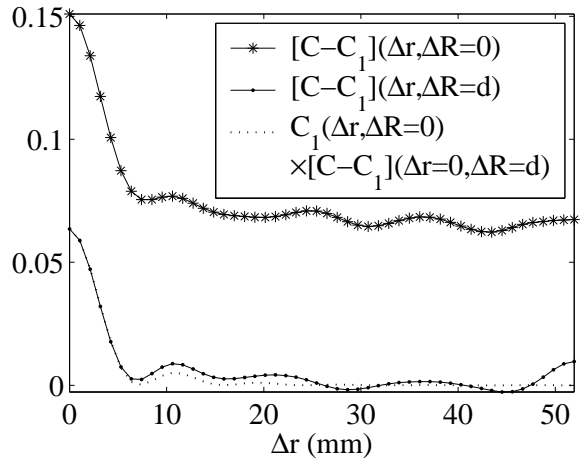}} \caption{plots of
$[C-C_1](\Delta r)$ for $ \Delta R= 0$ and 3 cm, and $C_1(\Delta
r)$ at $\Delta R = 0$.} \label{fig2}
\end{figure}

The spatial variations of $C(\Delta r,0)$ and $C_1(\Delta r,0)$
are shown in Fig.~\ref{fig1}. The $C_1$ contribution is directly
obtained by squaring the field correlation function, shown in the
inset of Fig.~\ref{fig1}. Subtracting $C_1$ from the full
intensity correlation function gives the difference $[C-C_1]$,
shown in Fig.~\ref{fig2} for a single source ($\Delta R=0$) and
for two sources separated by $\Delta R = d$. This difference
includes the terms beyond  the field factorization approximation.
\noindent Measurements of $C(0,\Delta R)$ and $C_1(0,\Delta R)$,
i.e. for a fixed detector and a scanning source, have also been
performed. Within experimental error, $C(\Delta r,0)$ and
$C(0,\Delta R)$ were found to be identical functions of their
respective arguments. The same is true for $C_1$. Similarly, we
find that plots of $[C-C_1]$ versus $\Delta R$ with $\Delta r = 0$
and $\Delta r = d$ are nearly the same as those shown in Fig. 2.
Thus $\Delta R$ and $\Delta r$ can be interchanged as required by
reciprocity.

\begin{figure}[h!]
\scalebox{.7}{\includegraphics{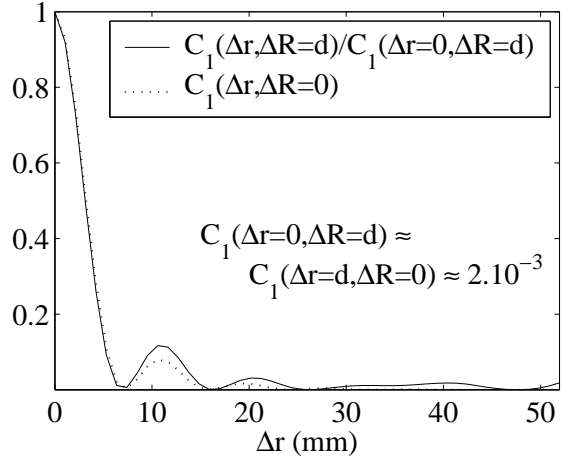}} \caption{Comparison
of $C_1(\Delta r)$, normalized by its value at $\Delta r=0$, for
$\Delta R=3$ cm and $C_1(\Delta r)$ at $\Delta R = 0$.}
\label{fig3}
\end{figure}
Measurements of $C_1(\Delta r, \Delta R)$ for $\Delta R = 0$ and
$\Delta R = d$ are presented in Fig.~\ref{fig3}. Within the noise
level of $10^{-4}$, the two functions have the same variation with
$\Delta r$, $C_1(\Delta r, d) = 2 \times 10^{-3}C_1(\Delta r, 0)$.
This numerical factor, $C_1(0,d)$, is equal to the value of
$C_1(d,0)$ within the uncertainty in $d$. This result, taken
together with the aforementioned symmetry with respect to
interchanging $\Delta r$ and $\Delta R$, suggests that $C_1$ can
be written as the product of two identical functions, $C_1(\Delta
r, \Delta R) = F(\Delta R)F(\Delta r)$.

We now examine $[C-C_1](\Delta r, \Delta R)$, which is do minated
by  $C_2$ in our sample. This function is seen in Fig.~\ref{fig2}
to fall to nearly one half its value when either $\Delta R$ or
$\Delta r$ increase beyond $\delta r$ when there is no
displacement of the other variable. This shows that $C_2$ is given
by the addition of two equal terms. The comparison of the
short-range variation of $[C -C_1]$ with $C_1$ in Fig.~\ref{fig2}
suggests that the additive form factors are identical to $F$, so
that the dominant contribution to $C-C_1$ is proportional to
$[F(\Delta R) + F(\Delta r)]$. This would imply that $C_2(\Delta
r)$ approach a constant value for $\Delta r > \delta r$, whereas
the measurement of $[C-C_1](\Delta r,d)$ is seen in
Fig.~\ref{fig2} to fall slightly with increasing displacement.
This could be the consequence of a slight departure from a
quasi-1D geometry at the output face of the sample. There, the
average intensity is slightly larger at the center than at the
edges since the wave can spread beyond the cross section of the
tube. Notwithstanding this effect, our experimental results
suggest that both $C_1$ and $C_2$ can be expressed in terms of a
single form factor $F(x)$, where $x$ stands for either $\Delta r$
or $\Delta R$. $C_1$ and $C_2$ contain, respectively, the product
and the sum of two form factors. In Fig.~\ref{fig2}, the
correlation function $[C-C_1](\Delta r,d)$ is seen to lie above
the dotted curve, which is proportional to $C_1$. This suggests a
constant contribution to $C$. For $\Delta r > 30$ mm, the
correlation function becomes negative, but here the noise becomes
larger than the signal because of the reduced number of pairs of
points with increasing $\Delta r$. Such a long-range correlation
for large values of $\Delta R$ may be part of $C_3$.

\begin{figure}[h!]
\scalebox{.7}{\includegraphics{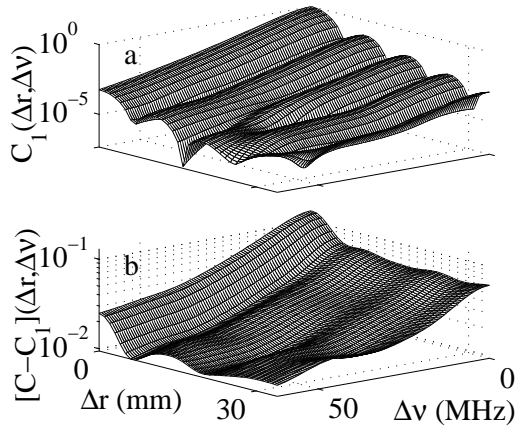}} \caption{Semilog
representation of the spatial and frequency dependence of $C_1$
(a) and $[C-C_1]$ (b) for $\Delta R=0$.} \label{fig4}
\end{figure}
The structure of the joint spatial and frequency dependences of
$C_1$ and $C_2$, is obtained from measurements of the correlation
functions $C_1(\Delta \nu,\Delta r)$ and $[C-C_1](\Delta
\nu,\Delta r)$ for $\Delta R=0$, shown in Figs.~\ref{fig4}a and
\ref{fig4}b, respectively. The semilog representations in
Fig.~\ref{fig4} show that, within the limits set by the noise
level, $C_i$ have the same frequency dependence for any $\Delta
r$, while $C_i$ have the same spatial dependence for any $\Delta
\nu$ for $i=1,2$. Thus their spatial and spectral variations for a
single source are given by $C_i(\Delta\nu,\Delta r)=A_i(\Delta
\nu)C_i(\Delta r)$. The noise level found in $C_1$ is low compared
to that in $C_2$ because the field correlation function, $F_E$, is
computed and then squared to obtain $C_1$, giving a signal to
noise ratio which is the square of that for the field correlation
function. The form of the intensity correlation function suggested
by experiment is borne out in diagrammatic calculations, which are
briefly summarized below.

The three terms in $C$ may be represented diagrammatically. The
diagram corresponding to the $C_1$ term describes two
non-interacting diffusons attached to pairs, $GG^\ast$, of
averaged Green's functions \cite{dutch,PniniBook}. This diagram
factorizes into a product of two field functions:
\be 
\av{E_\nu (\rr, \RR ) E_{\nu^\prime} (\rp , \Rp )} = \int d^3r_1
d^3 r_2 G_{\nu}(\rr, \vec{r}_1) G_{\nu^\prime}^\ast (\rp,
\vec{r}_1) T_{\nu\nu^\prime} (\vec{r}_1, \vec{r}_2 )
G_{\nu}(\vec{r}_2, \RR) G_{\nu^\prime}^\ast (\vec{r}_2, \Rp ) \ ,
\label{bethesalpeter} \ee where $T_{\nu\nu^\prime} (\vec{r}_1,
\vec{r}_2)$ denotes the diffusion ladder and integration is
performed over $\vec{r}_1$, $\vec{r}_2$ inside the tube. For the
quasi one-dimensional geometry, the diffuson is independent of its
transverse coordinates, whereas the Green's functions decay
rapidly on a scale of the mean free path. Taking $\RR =\Rp$ and
$\nu=\nu^\prime$, we obtain
\be 
\av{E_\nu (\rr, \RR ) E_{\nu} (\rp , \RR)} =
\left(\frac{4\pi}{\ell}\right) \int d^3r_1 G_{\nu}(\rr, \vec{r}_1)
G_{\nu}^\ast (\rp, \vec{r}_1)\av{I_\nu (\vec{r}_1, \RR)} \equiv
F_E(\Delta r)   \ , \label{form} \ee where $I_\nu (\rr, \RR)$ is
the intensity at $\rr$, normalized to its average value at the
output face of the tube. Thus, $C_1 (\Delta r, \Delta
R)=F_E^2(\Delta r) F_E^2 (\Delta R)\equiv F(\Delta R)F(\Delta r)$.

The diagrams corresponding to the $C_2$ and $C_3$ terms describe
two incoming and two outgoing diffusons which interact in the bulk
of the medium. In these diagrams, each pair $GG^\ast$ of external
Green's functions contributes a  spatial form factor $F_E$ as in
Eq.~(\ref{form}). These give
\bea 
C&=&C_1+C_2 +C_3 \nonumber \\
&=&  A_1(\Delta \nu,\alpha) F(\Delta R)F(\Delta r) +\frac{2}{3g}
A_2(\Delta \nu,\alpha) \left[ F(\Delta R) + F(\Delta
r) \right]  \nonumber \\
&+&\frac{2}{15g^2} A_3(\Delta \nu,\alpha) \left[ 1 +F(\Delta R) +
F(\Delta r) +F(\Delta R)F(\Delta r)\right] \ , \label{everybody}
\eea where $g$ is the average conductance of the sample. The
coefficients $A_i \  (i=1,2,3)$ depend on the absorption
coefficient $\alpha$ and the frequency shift $\Delta \nu$. The
structure of Eq.~(\ref{everybody}) is similar to that of the
correlation in transmission, obtained in the multichannel
formalism \cite{Feng88,mello,dutch}.  In the present case,
however, all terms are described by a single spatial form factor.
The field factorization term, $A_1(\Delta\nu=0,\alpha)$ is unity
and independent of absorption by definition, while
$A_2(\Delta\nu\!=\!0,\alpha)$ is given by \cite{kk,ps,brewer}:
\be
A_2 (\Delta\nu\!=\!0,\alpha) =  \frac{3}{16\alpha}\left[
\frac{\sinh 2\alpha -2\alpha (2 -\cosh 2\alpha)} {\sinh^2 \alpha}
\right] \ . \label{coeff} \ee The coefficient $A_3 (\alpha)$
depends weakly on $\alpha$ and its limiting values are $A_3(0)=1,
A_3(\infty)=15/16$.

For our samples, $g\approx 7$ and $\alpha\approx 3$ \cite{PRE00}.
From the measurement of $[C-C_1]$ at $\Delta r=0$, $\frac{2}{3g}
A_2 = 0.076$ with $A_2=0.87$. This is in agreement with
Eq.~(\ref{coeff}) and calculated corrections due to localization
effects \cite{kk,GGPS,ps,brewer,PniniBook}. Using the measured
$C_1(\Delta r)$ as the functional form $F(\Delta r)$, following
Eq.~(\ref{everybody}) and neglecting $C_3$, we obtain a good fit
of the spatial structure of the intensity correlation function, as
shown in Fig.~\ref{fig1}.

These considerations have applications to present efforts to
enhance the capacity of wireless communication by utilizing
multiple antennas to detect the multiply scattered field
\cite{Moustakas,PhysToday}. Antenna separation should be larger
than $\delta r$ and the number of statistically independent
antennas equals the inverse of the degree of long-range intensity
correlation, $3g/2$.

In conclusion, we have uncovered the connection between the field
and intensity correlation functions in the spatial structure of
the three contributions to $C$. In contrast to the case of angular
correlation \cite{Feng88}, intensity correlation can be expressed
in terms of a single form factor obtained from the field
correlation function. We have demonstrated the multiplicative
character of $C_1$ and the additive character of $C_2$.
Calculations predict a mixed character for $C_3$, which includes a
multiplicative, an additive, and a constant term of equal
amplitude. We observe the infinite-range component of $C_3$ in the
residual correlation when both the source and detector are
displaced by more than the correlation length. Determining the
proper breakup of $C$ into its components is of particular
importance when considering simultaneous variations in space, time
and frequency. Each term is a product of the corresponding $C_1$,
$C_2$ and $C_3$ correlation function in the appropriate variables.

We thank A.A. Chabanov for valuable discussions. Support from the
National Science Foundation (DMR 9973959), Army Research Office
(DAAD 190010362), the United States-Israel Binational Science
Foundation (BSF) and the Groupement de Recherche PRIMA are
gratefully acknowledged.



\begin{thebibliography}{}

\bibitem{Shapiro}
B. Shapiro, Phys. Rev. Lett. {\bf 57},  2168  (1986).

\bibitem{Stephen}
M.~J. Stephen and G. Cwilich, Phys. Rev. Lett {\bf 59},  285
(1987).

\bibitem{Feng88}S. Feng, C. Kane, P. A. Lee and A.
D. Stone, Phys. Rev. Lett. \textbf{61}, 834 (1988).

\bibitem{mello}
P.~A. Mello, E. Akkermans and B. Shapiro, Phys.\ Rev.\ lett.\ {\bf
61},459 (1988).


\bibitem{GenackPRL90}
A.~Z. Genack, N. Garcia, and W. Polkosnik, Phys. Rev. Lett. {\bf
65}, 2129 (1990).

\bibitem{kk}
E. Kogan and M. Kaveh, Phys.\ Rev.~B {45}, 1049 (1992).


\bibitem{dutch}
 M.C.W. Rossum and Th.M. Nieuwenhuizen, Rev.\ Mod.\ Phys.\ {\bf 71}, 313, 1999.

\bibitem{Shapiro99}
A point source embedded in the interior of a random medium exhibits
an additional $C_0$ correlation, which can be larger than $C_2$ in some
circumstances. B. Shapiro, Phys. Rev. Lett. \textbf{83}, 4733 (1999).

\bibitem{AltshulerAltshuler}
B.~L. Altshuler, V.~E. Kravtsov, and I.~V. Lerner,  in {\em
Mesoscopic Phenomena in Solids}, edited by B.~L. Altshuler, P.~A.
Lee, and R.~A. Webb (Noth-Holland, Amsterdam, 1991).



\bibitem{Laibowitz}
R.~A. Webb, S. Washburn, C.~P. Umbach, and R.~B. Laibowitz, Phys.
Rev. Lett. {\bf 54},  2696  (1985).




\bibitem{gang4}
E. Abrahams, P.~W. Anderson, D.~C. Licciardello, and T.~V. Ramakrishnan, Phys.\ Rev.\ Lett.\ {\bf 42},673 (1979).

\bibitem{Freund88}
I. Freund, M. Rosenbluh, and S. Feng, Phys. Rev.
Lett. {\bf 61}, 2328  (1988).

\bibitem{Li94}
J.~H. Li and A.~Z. Genack, Phys. Rev. E {\bf49}, 4530 (1994).

\bibitem{PRE00} P. Sebbah, R. Pnini and A. Z. Genack, Phys.\
Rev.~E, {\bf 62}, 7348 (2000).

\bibitem{Genack87}
A.~Z. Genack, Phys. Rev. Lett. {\bf 58}, 2043 (1987).

\bibitem{Garcia89}
N. Garcia and A.~Z. Genack, Phys. Rev. Lett. {\bf 63}, 1678 (1989).

\bibitem{Ad90}
M.~P. van Albada, J.~F. de~Boer, and A. Lagendijk, Phys. Rev.
Lett. {\bf 64}, 2787  (1990).

\bibitem{GGPS}
N. Garcia, A.~Z. Genack, R. Pnini, and B. Shapiro, Phys. Lett. A
{\bf 176}, 458 (1993).

\bibitem{Maret87}
G. Maret and P.~E. Wolf, Z. Phys.~B {\bf 65}, 409 (1987).

\bibitem{FSPRB97}F. Scheffold, W. H\"{a}rtl, G. Maret and  E. Matijevi\'{c},
Phys. Rev. B \textbf{56}, 10942 (1997).

\bibitem{FSPRL98}F. Scheffold and G. Maret, Phys. Rev. Lett. \textbf{\ 81}, 5800
(1999).

\bibitem{FreundEliyahou}
I. Freund and D. Eliyahu, Phys. Rev. A {\bf 45}, 6133 (1992).

\bibitem{ps}
R. Pnini and B. Shapiro, Phys.\ Lett.~A {\bf 157}, 265 (1991).

\bibitem{brewer}
P.~W. Brouwer, Phys.\ Rev.~B {\bf 57},10526 (1998).

\bibitem{PniniBook}
R. Pnini, in {\it Waves and Imaging through Complex Media}, ed. P.
Sebbah (Kluwer Academic Publishers, the Netherlands) p.391-412.

\bibitem{Moustakas}
A.~L. Moustakas, H.~U. Baranger, L. Balents, A.~M. Sengupta, and
S.~H. Simon, Science {\bf 287}, 287 (2000).

\bibitem{PhysToday}
S.~H. Simon, A.~L. Moustakas, M. Stoytchev, and H. Safar, Phys.
Today (Sept. 2001) 34.

\end{thebibliography}
\end{document}